# The relationship between the redox activity and electrochemical stability of solid electrolytes for solid-state batteries

Tammo Schwietert [#,1], Violetta Arszelewska [#,1], Chuang Yu[#,1], Chao Wang[1], Alexandros Vasileiadis[1], Niek J.J. de Klerk[1], Jart Hageman[1], Thomas Hupfer[2], Ingo Kerkamm[2], Yaolin Xu[1], Eveline van der Maas[1], Erik M. Kelder[1], Swapna Ganapathy[*,1] and Marnix Wagemaker[*,1]

[1]Storage of Electrochemical Energy, Faculty of Radiation Science and Technology, Delft University of Technology, Mekelweg 15, 2929JB, Delft, The Netherlands

[2] Robert Bosch GmbH, Corporate Sector Research and Advance Engineering, Robert-Bosch-Campus 1, 71272 Renningen

[#]Equally contributing authors

[*]Corresponding authors: s.ganapathy@tudelft.nl, m.wagemaker@tudelft.nl

**Abstract**

All-solid-state Li-ion batteries promise safer electrochemical energy storage with larger volumetric and gravimetric energy densities. A major concern is the limited electrochemical stability of solid electrolytes and related detrimental electrochemical reactions, especially because of our restricted understanding. Here we demonstrate for the argyrodite, garnet and NASICON type solid electrolytes, that the favourable decomposition pathway is indirect rather than direct, via (de)lithiated states of the solid electrolyte, into the thermodynamically stable decomposition products. The consequence is that the electrochemical stability window of the solid electrolyte is significantly larger than predicted for direct decomposition, rationalizing the observed stability window. The observed argyrodite metastable (de)lithiated solid electrolyte phases contribute to the (ir)reversible cycling capacity of all-solid-state batteries, in addition to the contribution of the decomposition products, comprehensively explaining solid electrolyte redox activity. The fundamental nature of the proposed mechanism suggests this is a key aspect for solid electrolytes in general, guiding interface and material design for all-solid-state batteries.

All-solid-state-batteries (ASSBs) are attracting ever increasing attention due to their high intrinsic safety, achieved by replacing the flammable and reactive liquid electrolyte by a solid electrolyte[1]. In addition, a higher energy density in ASSBs may be achieved through; (a) bipolar stacking of the electrodes, which reduces the weight of the non-active battery parts and (b) by potentially enabling the use of a Li-metal anode, which possesses the maximum theoretical Li capacity and lowest electrochemical potential (3860 mAhg$^{-1}$ and -3.04 V *vs.* SHE). First of all, the success of ASSBs relies on solid electrolytes with a high Li-ion conductivity[2-5]. A second prerequisite, is the electrochemical stability at the interfaces of the solid electrolyte with the electrode materials in the range of their working potentials. Any electrochemical decomposition of the solid electrolyte may lead to decomposition products with poor ionic conductivity that increase the internal battery resistance[2-4,6]. Third, ASSBs require mechanical stability as the changes in volume of the electrode materials upon (de)lithiation, as well as decomposition reactions at the electrode-electrolyte interface may lead to contact loss, also increasing the internal resistance and lowering the capacity[2-4].

Initially, for many solid electrolytes wide electrochemical stability windows were reported[4,7-10], which appeared in practice to be much more limited[4,11,12]. Evaluation of the electrochemical stability, based on differences in formation energies, indeed lead to much narrower stability windows[13,14], however, practical stability windows typically appear larger[4,11,12]. As a thermodynamic evaluation does not take into account kinetic barriers for decomposition reactions, which should be expected to play a critical role[12], the mechanisms of solid electrolyte decomposition are poorly understood, presenting one of the major challenges for ASSBs[2-4,11,12]. Another important aspect, directly related to this, is the potentially significant contribution of the typically Li-rich solid electrolytes through (de)lithiation reactions, either directly or indirectly[15]. In general, redox activity can be expected near the interface between the solid electrolyte and the electronically conductive components of the electrode (electrode active material and carbon additives[16]), but may also extend deep into the solid electrolyte through short range electron conductivity of the electrolyte itself[17]. Understanding the redox activity of solid electrolytes, and its correlation with the electrochemical stability window is thus of fundamental importance for the development of stable solid-solid interfaces in ASSBs.

Here, we demonstrate that the electrochemical stability window of the argyrodite $Li_6PS_5Cl$ solid electrolyte is determined by the solid electrolyte redox activity *i.e.* lithiation upon reduction of phosphorus and delithiation upon oxidation of sulfur, before decomposing into more stable products. As demonstrated by DFT simulations, this kinetically favourable indirect decomposition pathway effectively widens the electrochemical stability window, compared to direct decomposition into stable products, in excellent agreement with accurate electrochemical measurements. The (de)lithiated argyrodite phases are directly observed with XRD and solid state NMR, providing direct evidence of this indirect decomposition mechanism. As solid electrolytes are designed to provide fast ionic conduction, the indirect decomposition through (de)lithiation is proposed to be relevant for solid electrolytes in general, determining the practical electrochemical stability window. This is underlined by the agreement between the herein measured and the predicted indirect stability window for $Li_7La_3Zr_2O_{12}$ (LLZO) garnet type and $Li_{1.5}Al_{0.5}Ge_{1.5}(PO_4)_3$ (LAGP) NASICON type solid electrolytes. This mechanism establishes that not only the decomposition products, but also the solid electrolyte structure itself contribute to the reversible capacity in ASSBs, making the present findings highly relevant for the working and development of ASSBs.

**Electrochemical activity of the argyrodite Li$_6$PS$_5$Cl**

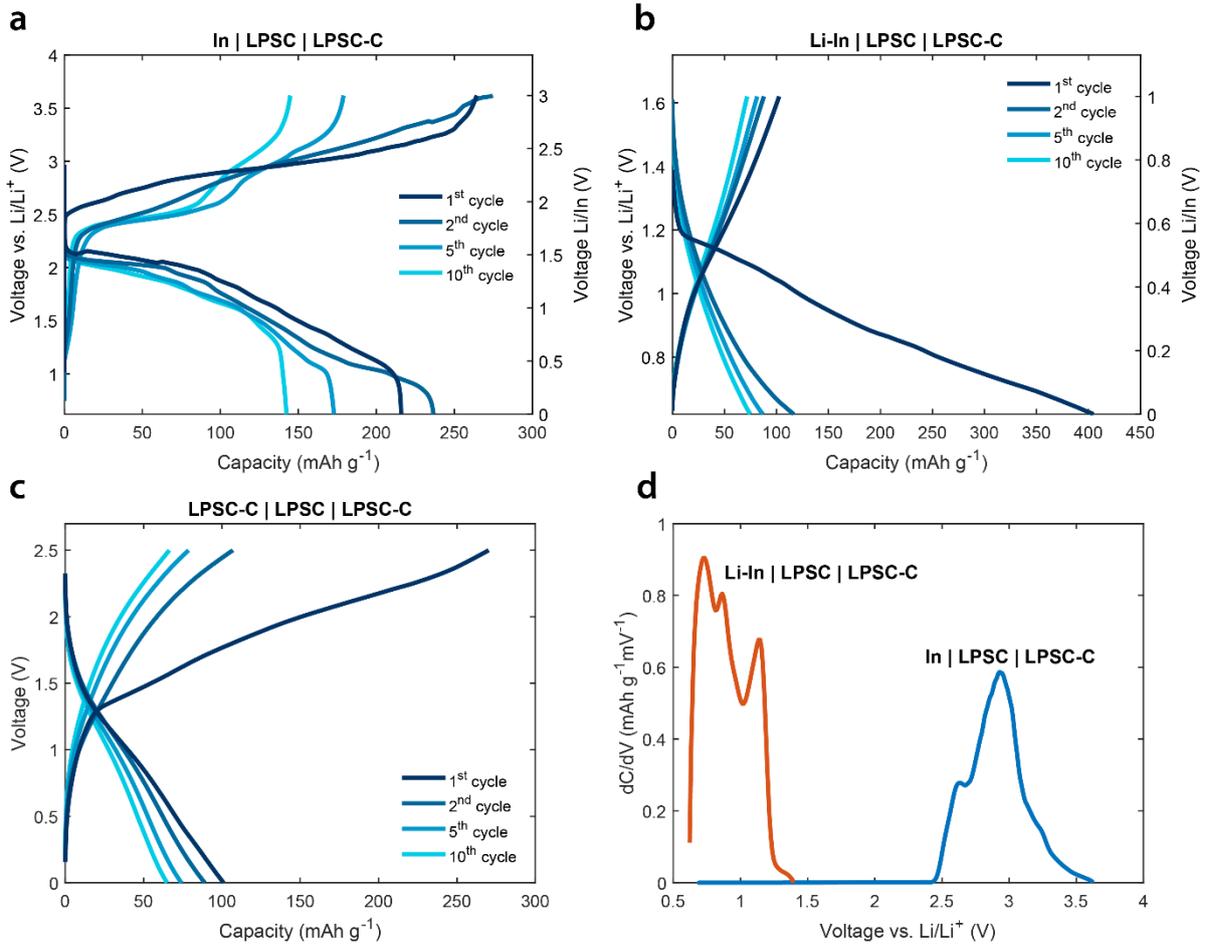

*Fig. 1: Voltage profiles and differential capacity curve of the Li$_6$PS$_5$Cl-C electrode. **a-c**, Voltage profile of the 1$^{st}$, 2$^{nd}$, 5$^{th}$ and 10$^{th}$ cycle of (**a**) In|LPSC|LPSC-C ASSB starting from charge, (**b**) Li-In|LPSC|LPSC-C ASSB starting from discharge and (**c**) LPSC-C|LPSC|LPSC-C, a one material (LPSC) ASSB. **d**, The differential capacity dC/dV curve of the In|LPSC|LPSC-C (blue) and the Li-In|LPSC|LPSC-C battery (orange) showing the first oxidation and first reduction of LPSC. Electrochemical activity is observed below 1.25 V and above 2.50 V vs. Li/Li$^+$, indicating an electrochemical stability window of 1.25 V.*

The electrochemical stability, especially for thiophosphate solid electrolytes, was shown to be significantly lower than initially expected[9,16,18-21], where the consequential decomposition reactions have had a large impact on the ASSB performance[10,20-23]. To investigate the electrochemical stability and the role of electrochemical reactions in solid electrolytes, the argyrodite $Li_6PS_5Cl$ (LPSC), introduced by Deiseroth et al.[24], is employed both as active material and solid electrolyte in ASSBs. To induce oxidation and reduction reactions of the solid electrolyte, carbon black and carbon nano-fibres are mixed in with the LPSC. The mixture is referred to as the LPSC-C electrode (for details see the methods section). To study the oxidation and reduction independently, while at the same time preventing the redox activity of the decomposition products from interfering with the decomposition itself, individual cells are prepared for the first oxidation and for the first reduction respectively. An In|LPSC|LPSC-C battery is cycled galvanostatically starting with oxidation, and a Li-In|LPSC|LPSC-C battery starting with reduction, the resulting voltage curves of which are shown in Figure 1a,b. Unless otherwise specified, all voltages are expressed vs. $Li/Li^+$. The galvanostatic current density used is very small (see methods section), especially because of the mixing with conductive carbon additives creating a very large interface with the solid electrolyte (>>1 $m^2$), which minimizes the contribution of overpotentials, thus approaching the thermodynamic potential. On galvanostatic oxidation, the LPSC-C electrode shows a voltage plateau at 2.5 V (Fig. 1a), reaching a total charge capacity of 264 mAh $g_{LPSC}^{-1}$ when charged to 3.63 V. On galvanostatic reduction, the LPSC-C electrode shows a voltage plateau at around 1.2 V (Fig. 1b), with a discharge capacity of 405 mAh $g_{LPSC}^{-1}$ when discharged to 0.63 V. The large partially reversible specific capacities demonstrate that LPSC can undergo severe oxidation and reduction reactions, and the low columbic efficiencies of 70 and 40% upon first oxidation and reduction, respectively, suggest the formation of a significant amount of decomposition products. The decreasing capacity of the initial cycles (Supplementary Fig. 1) indicates that these decomposition reactions increase the impedance. However, upon extended cycling, the reversible capacity remains relatively constant, which could indicate that the decomposition products are able to deliver reversible electrochemical activity, as already suggested for LPSC by Auvergnot et al.[25]. Since LPSC can undergo both oxidation and reduction reactions, it can be used to assemble a one-material-battery, similar to what was reported for the $Li_{10}GeP_2S_{12}$ solid electrolyte, for which the combination of decomposition products at the cathode and anode provided the reversible redox[15]. The assembled LPSC-C|LPSC|LPSC-C symmetric one-material-battery shows an initial charge capacity of 270 mAh $g_{LPSC}^{-1}$ shown in Figure 1c, which drops to 107 mAh $g_{LPSC}^{-1}$ in the second cycle. During the first charge the activity appears to set in at around 1.25 V, a direct indication of the practical electrochemical stability window. To evaluate the practical electrochemical stability window more accurately, the differential capacity is determined from the 1st charge of the In|LPSC|LPSC-C battery and from the 1st discharge of the Li-In|LPSC|LPSC battery, shown in Figure 1d. Indeed, a practical stability window of 1.25 V is obtained, much larger than that theoretically predicted (0.3 V)[13,14], and much smaller than initially reported (7 V)[9]. Additionally, the presence of more than one peak, both on reduction and oxidation, indicates subsequent redox activity. This raises the question; what reactions take place and how do these determine the observed electrochemical stability window?

**Evaluation of the redox activity of the argyrodite $Li_6PS_5Cl$ with DFT simulations**

Aiming for better understanding and prediction of practical electrochemical stability windows, and correlation with solid electrolyte redox activity, we evaluate the formation energies of all possible Li-vacancy configurations at different compositions of argyrodite $Li_xPS_5Cl$, within a single unit cell, similar to how the energetics of electrode materials are evaluated[26] (thus the simulations are performed on charge-neutral cells). This appears to be a realistic approach considering that the solid electrolyte is in contact with the conductive additives in a cathodic mixture, and therefore the solid electrolyte can function as an electrode material being oxidized and reduced. Argyrodite Li$_x$PS$_5$Cl crystallizes in the $F\bar{4}3m$ space group and at x = 6 has 50 % of the 48h crystallographic Li positions randomly occupied[6]. The starting structure of the argyrodite was obtained from literature, where a thorough investigation of the most stable configuration was performed taking into account the halogen disorder[27]. By calculating the energies of the symmetrically non-equivalent Li configurations, the most stable Li$_x$PS$_5$Cl configurations are obtained, from which the voltage at which these phases are formed can be determined (see computational methods).

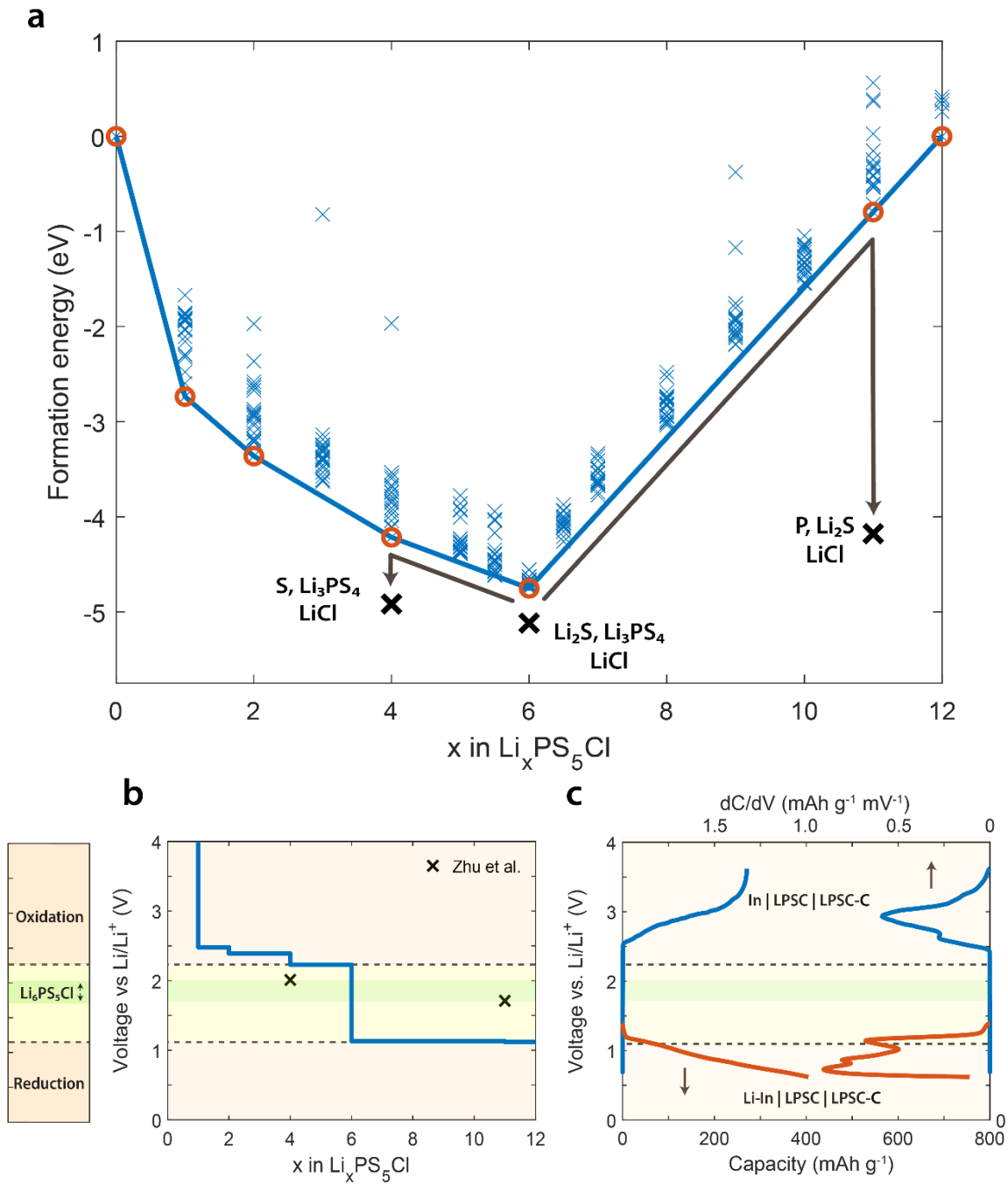

*Fig. 2: Formation energies of Li-vacancy configurations of $Li_xPS_5Cl$ and comparison of experimental and calculated voltage profiles. **a**, Formation energies per formula unit for all Li configurations within one unit cell, versus the composition x in $Li_xPS_5Cl$. The formation energy of the combination of $Li_3PS_4$, $Li_2S$ and LiCl is shown below the convex hull at x = 6. At x = 4 and x = 11 the formation energies of the decomposition products upon oxidation (S, $Li_3PS_4$, LiCl)*

*and upon reduction (P, Li$_2$S and LiCl) respectively, are shown, in line with the decomposition products previously reported[13]. **b**, Calculated theoretical voltage profile, vs. Li/Li$^+$, of Li$_x$PS$_5$Cl in the compositional range of 0 < x < 12. Reduction and oxidation are expected to occur at 1.08 V and 2.24 V vs. Li/Li$^+$, respectively. The black crosses indicate the voltages at which the argyrodite is expected to decompose, upon oxidation to S, Li$_3$PS$_4$, LiCl, and upon reduction to P, Li$_2$S and LiCl, previously reported[13]. **c**, First charge of the In|LPSC|LPSC-C (blue) and first discharge of the Li-In|LPSC|LPSC-C battery (orange) including the differential capacity from Figure 1d. Above 2.30 V vs. Li/Li$^+$ LPSC is oxidized, and below 1.25 V vs. Li/Li$^+$ LPSC is reduced. The legend compares the stability windows. Yellow: stability window of LPSC based on the oxidation and reduction potentials of Li$_4$PS$_5$Cl and Li$_{11}$PS$_5$Cl, respectively. Green: Predicted window (thermodynamic), based on the stability of the decomposition products for oxidation (S, Li$_3$PS$_4$, LiCl) and reduction (P, Li$_2$S and LiCl)[13,14].*

The resulting formation energies of the argyrodite Li$_x$PS$_5$Cl as a function of Li-composition are shown in Figure 2a, where the convex hull connects the most stable configurations. Upon oxidation and reduction of Li$_6$PS$_5$Cl the most stable compositions are Li$_4$PS$_5$Cl and Li$_{11}$PS$_5$Cl, respectively. Upon oxidation of argyrodite, sulfur is redox active (S$^{-2}$ → S$^0$ + 2e$^-$) whereas upon reduction, phosphorous is redox active (P$^{5+}$ → P$^0$ – 5e$^-$). Also indicated in Figure 2a are the energies of the thermodynamically more stable combinations of Li$_3$PS$_4$, Li$_2$S and LiCl species, and the most stable decomposition products of the oxidized and reduced argyrodite. Clearly, a delithiated (oxidized) argyrodite (Li$_4$PS$_5$Cl) is much less stable than the combination of Li$_3$PS$_4$, S and LiCl, as previously predicted[13,14], which are therefore the expected decomposition products on oxidation[13,14]. Similarly, lithiated (reduced) argyrodite (Li$_{11}$PS$_5$Cl) is much less stable than the combination of P, Li$_2$S and LiCl[13,14], which are therefore the expected decomposition products on reduction[13,14].

The average voltages, calculated from the convex hull, as a function of Li composition x in $Li_xPS_5Cl$ are shown in Figure 2b. From the theoretical voltage curve it is expected that the argyrodite LPSC delithiates (oxidizes) at 2.24 V and lithiates (reduces) at 1.08 V. This indicates that, if the decomposition reactions for oxidation and reduction are determined by the stability of the $Li_4PS_5Cl$ and $Li_{11}PS_5Cl$ species respectively, an electrochemical stability window of 1.16 V is expected. Also indicated is the much narrower electrochemical stability window, approximately 0.3 V wide, based on direct decomposition into $Li_3PS_4$, S and LiCl (oxidation) and into $Li_3PS_4$, $Li_2S$ and LiCl (reduction), in line with previous DFT calculations[13,14]. Which stability window applies, depends on the activation barriers to these decomposition routes. Based on the high Li-ion conductivity of the argyrodite, indicating low kinetic barriers for changes in the Li-composition, we propose that the decomposition occurs indirectly, via the lithiated and delithiated compositions of argyrodite ($Li_4PS_5Cl$ and $Li_{11}PS_5Cl$), rather than directly into the decomposition products. Upon argyrodite oxidation and reduction, first $Li_4PS_5Cl$ and $Li_{11}PS_5Cl$ would form, which are most likely unstable as evaluated below, providing a facile reaction pathway towards the formation of the more stable decomposition products as indicated by the solid black arrows in Figure 2a.

The experimental voltage curves obtained on oxidation and reduction of the argyrodite, including their differential capacity, are shown for comparison in Figure 2c. Remarkable agreement is found between the predicted electrochemical stability window of 1.16 V (Fig. 2b) and the experimentally observed window (Fig. 2c), supporting the present hypothesis that the argyrodite stability is determined by its redox activity upon (de)lithiation. The formation of decomposition products can be expected to increase the impedance depending on their location in the electrodes, which is most likely the origin of the broadening of oxidation and reduction peaks in the differential capacity shown in Fig. 2c. The offset between the measured and predicted stability window is most likely a result of the lower voltages predicted by DFT calculations[28]. Based on this, we propose that the first oxidation peak in the differential capacity, shown in Figure 1d and Figure 2c, is associated with the decomposition of LPSC at around 2.24 V via $Li_4PS_5Cl$ into $Li_3PS_4$, S and LiCl, and the first reduction peak in the differential capacity with the decomposition of LPSC at around 1.08 V via $Li_{11}PS_5Cl$ into P, $Li_2S$ and LiCl.

Upon further oxidation, after the formation of $Li_3PS_4$ via $Li_4PS_5Cl$, thermodynamic evaluation predicts the formation of $P_2S_5$ at 2.3 V [13]. Further reduction, after formation of P via $Li_{11}PS_5Cl$, thermodynamic evaluations predicts the formation of $Li_3P$, should display several potentials starting from 1.3 V down to approximately 0.87 V, the latter representing 67% of the reduction capacity (LiP to $Li_3P$) (see Supplementary Table 1, consistent with previous DFT predictions[29]). This provides a complete predicted oxidation and reduction potential pathway, via the solid electrolyte to the redox activity of the decomposition products as illustrated by Supplementary Fig. 2. For reduction this is consistent with the observed reduction activity measured at around 0.8 V in Fig. 2d, which is consistent with the known reduction potentials associated with the lithiation of phosphorus[30]. However, upon oxidation of the expected $Li_3PS_4$ decomposition product, a peak in the differential capacity is observed around 2.9 V, not in agreement with the predicted $P_2S_5$ formation at 2.3 V. As discussed below, formation of $P_2S_7^{4-}$ is observed consistent with the P-S-P bridging polyhedral suggested by XPS[19]. Moreover, $Li_3PS_4$ has been observed to oxidize at 2.9 V towards of $P_2S_7^{4-}$ [31] consistent with our observed oxidation activity in Fig. 1d and 2b. We anticipate that to predict the oxidation of $Li_3PS_4$ to $P_2S_7^{4-}$ at 2.9 V requires a detailed DFT redox activity analysis as done here for LSPC.

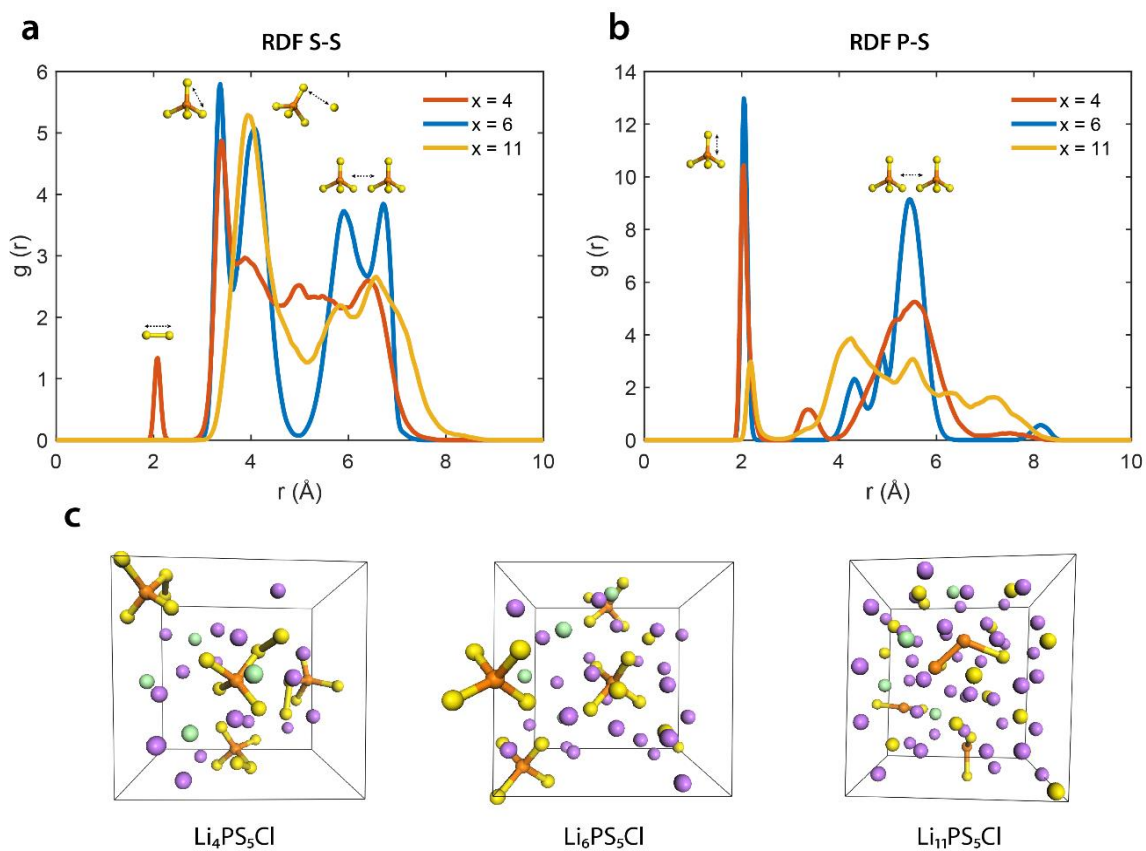

**Fig. 3: Molecular dynamics simulations of $Li_4PS_5Cl$, $Li_6PS_5Cl$ and $Li_{11}PS_5Cl$ a,** Radial distribution function (RDF) of the S-S bonds in (de)lithiated $Li_xPS_5Cl$ for x = 4, 6, and 11 during a 400 K DFT-MD simulation. During delithiation an increase in S-S bonds is seen around 2.1 Å, indicating that the formation of S-S bonds originates from the oxidation of S in the argyrodite. On top of the peaks, bonds at corresponding radii are displayed. **b,** Radial distribution function (RDF) of the P-S bonds of (de)lithiated $Li_xPS_5Cl$ for x = 4, 6 and 11 during a 400 K DFT-MD simulation, showing a decrease of P-S bonds in $PS_4$ units, indicating the reduction of P occurs in the argyrodite. **c,** Relaxed structures of $Li_xPS_5Cl$ for x = 4, 6 and 11 after a 400 K DFT-MD simulation. The violet, orange, yellow and green spheres indicate lithium, phosphorous, sulfur, and chlorine respectively. For the $Li_4PS_5Cl$ structure S atoms at the 4a and 4c positions form S-S bonds with $PS_4$ groups, while for $Li_{11}PS_5Cl$ P-S bonds break upon reduction of P.

To evaluate the kinetic stability of the delithiated ($Li_4PS_5Cl$) and lithiated ($Li_{11}PS_5Cl$) phases, DFT based molecular dynamics (MD) simulations are performed. It is important to realize that the timescale at which these structural transformations can be evaluated is very limited, up to 100 picoseconds at present, and therefore sluggish transformations fall outside the scope of this evaluation. The radial distribution functions and corresponding (de)lithiated structures of $Li_xPS_5Cl$ with = 4, 6, and 11 after the MD simulations are shown in Figure 3. In the delithiated phase, $Li_4PS_5Cl$, the S atoms at the 4a and 4c positions bond to the $PS_4$ groups, demonstrated by a decrease in the intensity at $r$ = 4.1 Å and an increase of intensity at $r$ = 2.1 Å (Fig. 3a), consistent with experimental observations where S bonds $PS_4$ groups[19,20]. Because the $PS_x$ units move relative to each other in $Li_4PS_5Cl$, peak broadening occurs for r > 5 Å. For the lithiated phase, $Li_{11}PS_5Cl$, a drop in intensity is observed at r = 2.1 Å (Fig. 3b), consistent with breaking P-S bonds in the $PS_4$ groups. This is expected because the P atoms can compensate for the change in valence as a consequence of the lithiation. The peak broadening of the lithiated structures indicates disorder in the S positions, resulting in a less defined structure in the simulated cell. In contrast, but as expected, no structural changes are observed for LPSC, reflecting its metastability versus $Li_3PS_4$, $Li_2S$ and LiCl (see also Fig. 2a). The MD simulations indicate that the $Li_4PS_5Cl$ and $Li_{11}PS_5Cl$ compositions are extremely unstable, having very low activation barriers towards decomposition. Their instability suggests that these compositions will only occur locally in the material, rapidly initiating local decomposition, which will nevertheless require the associated oxidation or reduction potential predicted by the convex hull shown in Fig. 2b. This supports the presently proposed indirect decomposition reaction, via the facile oxidation and reduction of the argyrodite towards the $Li_4PS_5Cl$ and $Li_{11}PS_5Cl$ phases, respectively, further decomposing into the oxidative ($Li_3PS_4$, S and LiCl) and reductive (P, $Li_2S$ and LiCl) decomposition products.

**Analysis of the argyrodite Li$_6$PS$_5$Cl decomposition products using XRD and $^6$Li and $^{31}$P solid-state NMR**

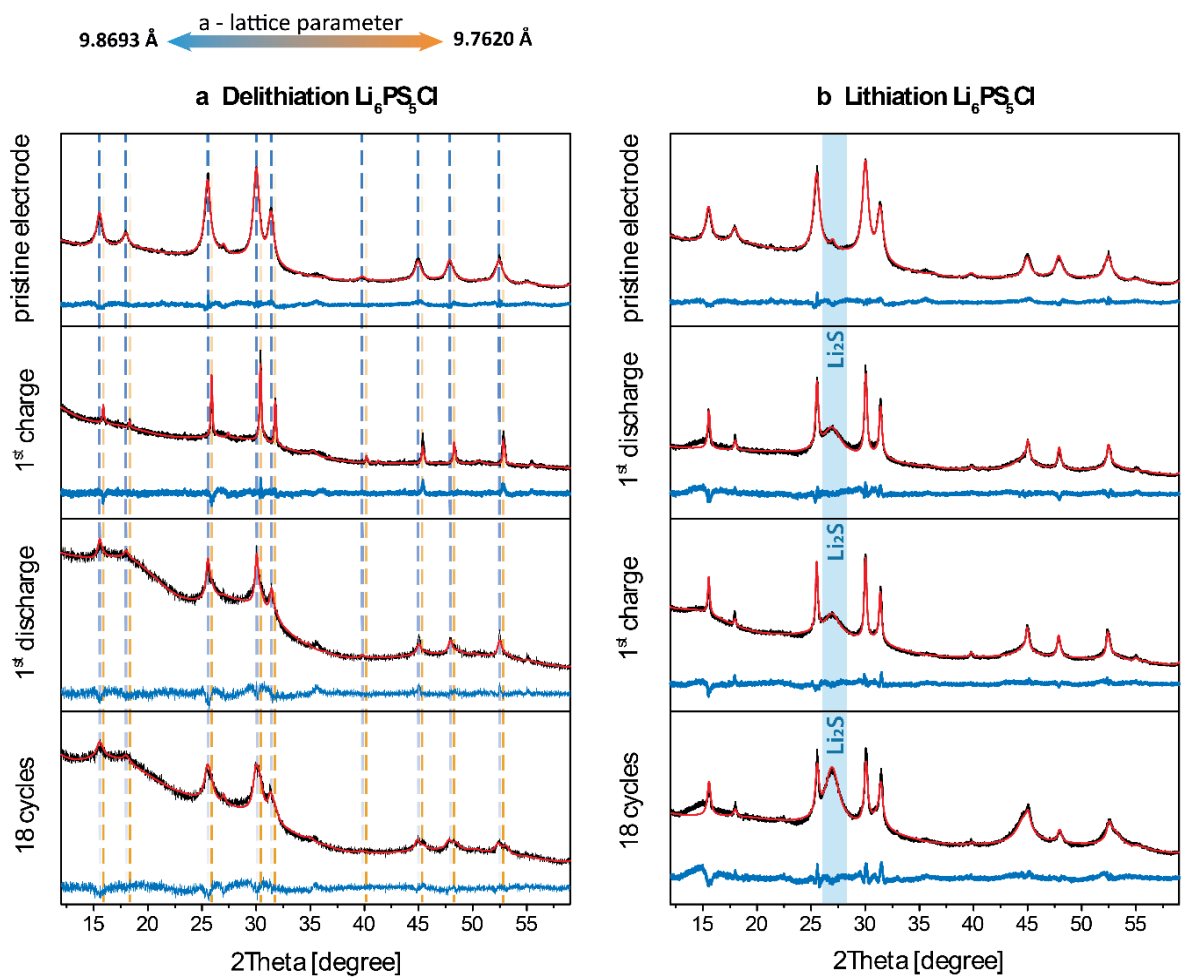

*Fig. 4: XRD patterns and fits of the LPSC-C electrodes before and after cycling.* All the patterns are fit with the Rietveld method as implemented in GSAS$^{32}$, the resulting structural parameters of which are provided in Supplementary Table 2-7. *a*, XRD patterns for the In|LPSC|LPSC-C battery after 1$^{st}$ charge to 3.63 V vs Li/Li$^+$, after subsequent discharge to 0.63 V vs Li/Li$^+$ and after 18 full cycles. *b*, XRD patterns for the Li-In|LPSC|LPSC-C battery after 1$^{st}$ discharge to 0.63 V vs Li/Li$^+$, after subsequent charge to 1.63 V vs Li/Li$^+$ and after 18 full cycles.

To monitor the structural changes of LPSC-C electrodes, XRD measurements are performed at different stages during the cycling of both the In|LPSC|LPSC-C and Li-In|LPSC|LPSC-C batteries as shown in Figure 4, following the cycling shown in Figure 1a,b. After the 1$^{st}$ half cycle, both on oxidation and reduction, a decrease in peak width is observed, indicating an increase in average crystallite size. The average crystallite size increases from 13 nm to 80 nm on delithiation (Fig. 4a) and to 41 nm on lithiation (Fig. 4b) respectively. An increase in average particle size can be rationalized by the preferential decomposition of smaller particles, most likely due to their large surface areas and resulting shorter electronic pathways for oxidation and reduction. This implies that electronic transport occurs through the argyrodite solid over tens of nanometers (the size of argyrodite particles). Upon subsequent cycling, the argyrodite XRD peaks widen, which may indicate partial decomposition of larger particles as well as a distribution of argyrodite lattice parameters as discussed below.

During the oxidation (delithiation) of the LPSC-C electrode to 3.63 V, the LPSC peak positions shift (Fig. 4a), corresponding to a decrease in the average cubic lattice parameter from 9.87 Å to 9.76 Å. This can be attributed to the partial delithiation of the LPSC phase, consistent with the lattice volume changes predicted by DFT for the compositional range $6 \geq x \geq 4$ for $Li_xPS_5Cl$ (Supplementary Fig. 3). Interestingly, several argyrodite compositions between $Li_4PS_5Cl$ and $Li_{11}PS_5Cl$ are located slightly above the convex hull (only 7.5 meV per atom for $Li_5PS_5Cl$) as seen in Figure 2a. Based on the convex hull in Fig. 2a, these metastable phases $6 \geq x \geq 4$ should disproportionate into $Li_4PS_5Cl$ (which would decompose immediately) and $Li_6PS_5Cl$. However, in reality, the system will not be in thermodynamic equilibrium as some parts of the electrodes are, or can become, more isolated through poor ionic and/or electronic contact. This makes it reasonable to suggest that parts of the electrode can be captured in 6

≥ x ≥ 4 metastable phases (which are kinetically more stable as compared to the Li$_4$PS$_5$Cl and Li$_{11}$PS$_5$Cl phases). Importantly, the presence of these phases in the composition range 6 ≥ x ≥ 4, provides experimental support for the proposed indirect decomposition mechanism via (de)lithiation of the solid electrolyte. After subsequent reduction, hence after one complete charge-discharge cycle, two different phases of argyrodite appear to be present, indicated by the dashed lines in Figure 4a. The 2θ position of the first phase (blue line) shifts back to the pristine argyrodite position, indicating that at least a partially reversible (de)lithiation of LPSC occurs. The second phase (orange line) remains at the position representing the delithiated argyrodite phases, the amount of which appears to grow upon cycling, indicating an increasing amount of oxidized argyrodite phases are formed upon cycling. The total amount of crystalline argyrodite decreases as indicated by the increasing background that appears over cycling, indicating the concomitant formation of amorphous sulfide and phosphorous sulfide decomposition products.

During the first reduction (lithiation) of the LPSC-C electrode to 0.63 V, the XRD patterns (Fig. 4b) do not display an obvious peak shift, as would be expected for the lithiated phases of argyrodite (Supplementary Fig. 3). A growing peak at around 27° reflects the formation of the Li$_2$S phase, consistent with the predicted decomposition reaction of lithiated (reduced) argyrodite (*Li$_{11}$PS$_5$Cl → P + 5 Li$_2$S + LiCl*). The amount of the Li$_2$S phase that is formed increases dramatically as a function of cycle number, also indicating the continued decomposition of the argyrodite for low potential cycling.

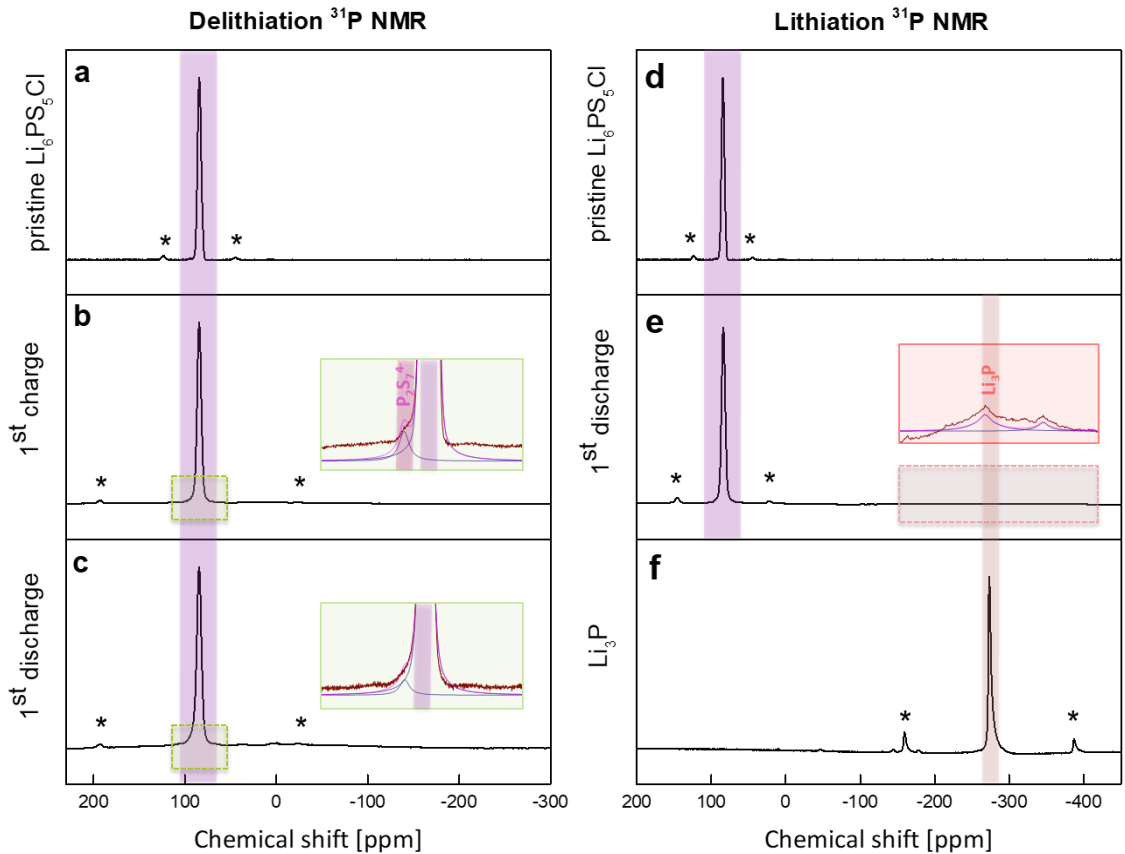

***Fig. 5: Solid-state $^{31}$P NMR spectra of pristine, oxidized and reduced LPSC-C. a-f,*** *$^{31}$P MAS NMR of oxidative (**a–c**) and reductive (**d–f**) activity of LPSC in In|LPSC|LPSC-C and Li-In|LPSC|LPSC-C ASSBs respectively. After the first charge, $P_2S_7^{4-}$ is found in the cathodic mixture, and after first discharge $Li_3P$ is present in the anodic mixture, in agreement with DFT and thermodynamic predictions. Spinning sidebands are indicated with an asterisk.*

Complementary to the XRD measurements, solid-state $^6$Li and $^{31}$P MAS NMR measurements are performed to analyze the decomposition products formed upon cycling. For pristine argyrodite, the $^{31}$P resonance at 85 ppm, shown in Figure 5a,d, can be assigned to the P environment in the $PS_4$ tetrahedral units[33]. After the 1$^{st}$ oxidation (delithiation) to 3.63 V of the LPSC-C electrode, an additional shoulder is observed at 95 ppm (Fig. 5b) which can be assigned to the $^{31}$P environment of $P_2S_7^{4-}$ species[34-36]. This indicates the formation of S-S bonds between PS$_4$ tetrahedral units (P-S-S-P), which undergoes a disproportionation reaction leading to the formation of P$_2$S$_7^{4-}$ and S$^0$, with P-S-P bridging polyhedra[19]. Upon 1$^{st}$ oxidation, the $^6$Li NMR spectrum (Supplementary Fig. 4b) shows the formation of an additional shoulder at around -1.1 ppm consistent with the formation of LiCl[37]. This supports the decomposition products observed by XPS[22,25], and is also in line with the MD simulations that indicate the bonding of S to PS$_4$ units. Note that the oxidation to Li$_3$PS$_4$, S and LiCl is proposed at 2.24 V, via the intermediate formation of Li$_4$PS$_5$Cl, whereas at around 2.9 V the oxidation towards P$_2$S$_7^{-4}$ and S$^0$ can be expected (Fig. 2b), all due to the S/S$^{-2}$ redox, represented by the first and second oxidation peaks of the differential capacity (Fig. 1d). The line broadening of the $^{31}$P and $^6$Li resonances of LPSC may originate from a distribution in bond angles and Li-deficient phases observed with XRD (Fig. 4a). After a full cycle i.e. 1$^{st}$ oxidation to 3.63 V followed by reduction to 0.63 V, the intensity of the amount of P$_2$S$_7^{4-}$ decreases, whereas in the $^6$Li NMR spectrum, a new Li-environment appears at 0.44 ppm which can be assigned to Li$_3$PS$_4$ (Supplementary Fig. 4c,d). This indicates that the P-S-P bridges connecting the PS$_4$ units, forming upon oxidation, break upon reduction transforming them back to isolated PS$_4$ units, similar to what was reported for the Li$_3$PS$_4$ electrolyte[19,20,38].

Upon the 1$^{st}$ reduction (lithiation) to 0.63 V of the LPSC-C electrode, a new $^{31}$P environment appears at -220 ppm (Fig. 5e) which can be assigned to Li$_3$P (Fig. 5f). The $^6$Li NMR spectrum (Supplementary Fig. 4f), shows the appearance a Li chemical environment very similar to that of Li in the argyrodite. Although the $^6$Li chemical shift of this environment is close to that of Li$_2$PS$_3$ (Supplementary Fig. 5), the associated phosphorus environment at 109 ppm is not observed in Figure 5e. We suggest that this Li environment may represent disordered lithiated argyrodite phases, which are suggested to form as metastable phases, occurring just above the convex hull in Figure 2a. Also, an additional peak appears at 2.3 ppm in the $^6$Li spectrum (Supplementary Fig. 4f), which can be assigned to the formation of Li$_2$S, consistent with the XRD pattern in Figure 4b. After a full cycle i.e. 1$^{st}$ reduction to 0.63 V followed by oxidation to 1.63 V, Li$_3$P disappears (Supplementary Fig. 6), indicating that in this voltage range phosphorous is redox active, reversibly transforming Li$_3$PS$_4$ to Li$_3$P. The observed formation of Li$_3$P and Li$_2$S in the LPSC-C electrodes reduced to 0.63 V, is consistent with XPS observations showing the formation of Li$_3$P, Li$_2$S and LiCl species, at the interface of LPSC with Li-metal[21]. The formation of P, Li$_2$S and LiCl, through the decomposition of the intermediate Li$_{11}$PS$_5$Cl, is expected to occur at 1.08 V, and further reduction up to 0.63 V will result in the formation of Li$_3$P at around 0.8 V[30] as observed (Fig. 2d) and predicted (Fig. 2b).

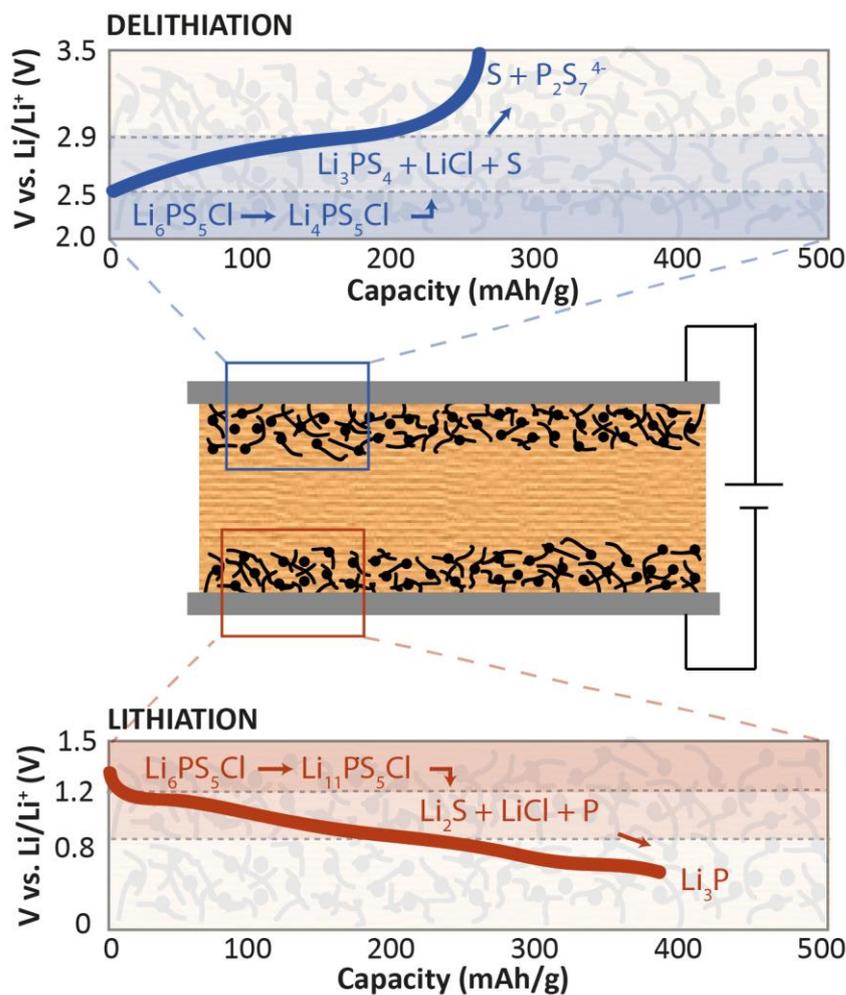

***Fig. 6: Schematic of the electrochemical activity of Li$_6$PS$_5$Cl upon oxidation (delithiation) and reduction (lithiation)*** *The electrochemical stability window is determined by the oxidation and reduction potentials of Li$_4$PS$_5$Cl (effectively S/S$^{2-}$ redox) and Li$_{11}$PS$_5$Cl (effectively P/P$^{5+}$ redox), respectively, here shown schematically, where the activation barriers to forming these compositions are expected to be very low based on the high Li-ion conductivity. These argyrodite compositions are highly unstable, having low activation barriers, resulting in rapid decomposition into the thermodynamically more stable products. The decomposition products provide reversible capacity upon ASSB cycling. Additionally, metastable oxidized and reduced*

*argyrodite phases form i.e. $Li_{4<x<6}PS_5Cl$ and $Li_{6<x<11}PS_5Cl$ respectively, that provide additional reversible capacity upon cycling solid state batteries.*

The proposed indirect oxidative and reductive decomposition of the argyrodite LPSC solid electrolyte, via the unstable $Li_4PS_5Cl$ ($S/S^{2-}$ redox) at 2.24 V and unstable $Li_{11}PS_5Cl$ (through the $P/P^{5+}$ redox) at 1.08 V, is schematically shown in Figure 6. These redox potentials of the argyrodite solid electrolyte determine the practical electrochemical stability window, as expressed by the first oxidation and reduction reactions observed in the cycling (Fig. 1a,b), and in the differential capacity (Fig. 1d), consistent with the predicted redox activity (Fig. 2). These unstable argyrodite phases rapidly decompose into the expected stable $Li_3PS_4$, S and LiCl species after oxidation, and P, $Li_2S$, and LiCl species after reduction. These decompose upon further oxidation and reduction to $P_2S_7^{-4}$ and $S^0$ at 2.9 V[31] and $Li_3P$ around 0.8 V[13,14] respectively, as observed by XPS[19,20] and the present XRD and NMR analysis. XRD and NMR also demonstrate the presence of metastable (de)lithiated argyrodite phases. This provides strong support for the proposed kinetically most favourable decomposition route, via the redox activity of the argyrodite solid electrolyte, thereby determining the electrochemical stability window. Both the redox activity of the solid electrolyte and of the decomposition products are responsible for the observed cycling capacity at anodic and cathodic potentials. In ASSBs this implies that both contributions of the solid electrolyte, will add to the cycling capacity based on the active electrode materials and the specified potential ranges. Moreover, the poor ionic conductivity of the decomposition products, especially S, $Li_2S$ and LiCl, as well as the change in volume can be expected to be responsible for the large increase in interfacial resistance upon cycling[21,39,40]. In addition to the observed decomposition reactions, specific active materials can result in additional decomposition reactions, for instance $Ni_3S_4$ upon cycling LPCS in combination with a NCM622 (LiNi$_{0.6}$Co$_{0.2}$Mn$_{0.2}$O$_2$) cathode (Supplementary Fig. 7).

The practical reversible and irreversible electrochemical activity, of the electrolyte in ASSBs, either originating from decomposition reactions or from the solid electrolyte itself, depends to a large extent on the electronic contact with the current collector which in turn depends on the electrode morphology. The redox activity of the solid electrolyte is expected at its contact areas with the electronically conductive cathode material and conductive carbon additive, but may also extend deep into the solid electrolyte itself, as solid electrolytes may be able to conduct electrons over small distances, as demonstrated for instance for Li$_3$PS$_4$[17]. In line with that the present detailed XRD analysis shows that decomposition reactions are not limited to the near interface area, as complete, tens of nanometer large, solid electrolyte particles decompose.

The fundamental nature of the present decomposition mechanism, through the (de)lithiation of the solid electrolyte, suggests that it is highly relevant for solid electrolytes in general. To support this, the Li insertion/extraction potentials are determined for two different type of solid electrolytes, garnet LLZO and NASICON LAGP. The convex hull shown in Fig. 7a indicates that the (delithiation) oxidation of Li$_7$La$_3$Zr$_2$O$_{12}$ results in an average calculated voltage of 3.54 V as shown in Fig. 7b, which is significantly larger compared to the direct decomposition at 2.91 V based on the stability of the predicted decomposition products Li$_2$O$_2$, La$_2$O$_3$ and Li$_6$Zr$_2$O$_7$[13]. It is unlikely that oxidation will proceed to Li$_1$La$_3$Zr$_2$O$_{12}$ as suggested by the convex hull, because several compositions between x=7 and x=6 (in Li$_x$La$_3$Zr$_2$O$_{12}$) are located marginally above the convex hull. This suggests that in the presence of slightly higher potentials (>3.54 V), oxidation will lead to indirect decomposition via x=6.5, towards the predicted stable decomposition products $Li_2O_2$, $La_2O_3$ and $Li_6Zr_2O_7$[13]. This is confirmed by extremely slow first galvanostatic oxidation of LLZO, shown in Fig. 7c, demonstrating that LLZO oxidation indeed occurs above 3.54 V. For LAGP, the convex hull in Fig. 7d predicts that reduction (lithiation) of $Li_{1.5}Al_{0.5}Ge_{1.5}(PO_4)_3$ occurs at 2.31 V which is lower than direct decomposition at 2.70 V based on the stability of the predicted decomposition products Ge, $GeO_2$, $Li_4P_2O_7$ and $AlPO_4$[13]. Consistently, a small but distinguishable reduction peak is observed at 2.4 V upon extremely slow first reduction of LAGP. We anticipate that reduction will not go through the highly lithiated $Li_7Al_{0.5}Ge_{1.5}(PO_4)_3$ composition, but indirectly through lower lithiated states. How exactly this proceeds is a subject of further study, however, the main message here is that the LAGP stability is determined by the initial (lithiation) reduction potential. These results support that the proposed indirect, kinetically favorable decomposition, via the (de)lithiation of the solid electrolyte is a general mechanism, in practice widening the solid electrolyte stability window. Notably, the reduction of LLZO and the oxidation of LAGP are not considered at present because both the indirect and direct reduction result in practically the same potential, making it impossible to discriminate between the two different mechanisms.

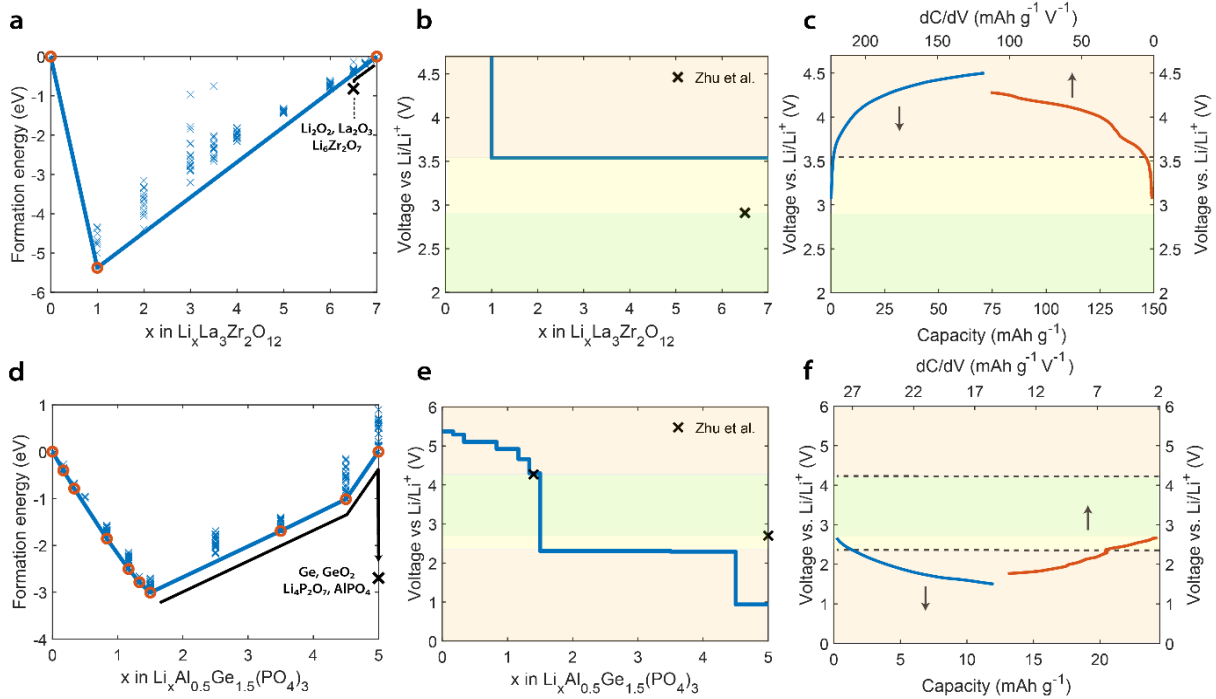

**Fig. 7: Formation energies of Li-vacancy configurations of garnet $Li_xLa_3Zr_2O_{12}$ (LLZO) and NASICON $Li_{1.5}Al_{0.5}Ge_{1.5}(PO_4)_3$ (LAGP) solid electrolytes and comparison of experimental and calculated oxidation potentials a,d** Formation energies per formula unit of $Li_xLa_3Zr_2O_{12}$ for $0 \leq x \leq 7$ and for $Li_xAl_{0.5}Ge_{1.5}(PO_4)_3$ $0 \leq x \leq 5$. The formation energy of the decomposition products expected for thermal equilibrium are indicated with a black cross. **b,e** The calculated voltage based on the convex hull of $Li_xLa_3Zr_2O_{12}$ and $Li_xAl_{0.5}Ge_{1.5}(PO_4)_3$. The blue line represents the redox potentials of the solid electrolytes, and the potentials for direct oxidation/reduction into the decomposition products are indicated with a black cross[13]. The green area indicates the stability window assuming direct decomposition, defined by the black cross, and yellow the extended stability window based on the presently proposed indirect decomposition via (de)lithiation of the solid electrolyte. **c,f** Experimental voltage curve and differential capacity upon first oxidation of a Li | Liquid Electrolyte | LLZO-C battery and first reduction of a Li | Liquid Electrolyte | LAGP-C battery. The differential capacity shows that oxidation of LLZO occurs around 3.6 V and reduction of LAGP occurs around 2.4 V, both in good agreement with *the predicted stability window based on the indirect decomposition via (de)lithiation of the solid electrolytes. The specific capacities are calculated based on the weight of LLZO and LAGP respectively.*

As solid electrolytes are designed for high ionic conductivity, the activation energies for oxidation and reduction reactions, associated with delithiation and lithiation respectively, can be expected to be small. The resulting metastable solid electrolyte compositions provide a kinetically facile reaction intermediate, providing an indirect pathway towards the more stable solid electrolyte decomposition products. As a consequence, the electrochemical stability window is determined by the solid electrolyte oxidation and reduction potentials (S and P redox for argyrodite and several other thiophosphate based solid electrolytes, Oxygen and Zr redox for LLZO and Oxygen and P redox for LAGP), and not by the stability of the most stable solid electrolyte decomposition products. The consequence of this indirect thermodynamic pathway, is that the electrochemical stability window is generally wider than that based on only on the stability of the decomposition products. Based on this mechanism, the design of stable solid electrolytes and their interfaces should focus on maximizing (de)lithiation redox potentials of the solid electrolytes. The demonstrated relation between the solid electrolyte electrochemical stability window and the redox reactions of the electrolyte, are decisive for the performance of solid state batteries and provide understanding that will contribute to the design electrolyte-electrode interfaces in ASSBs.

**Materials and Methods**

**Synthesis:**

Argyrodite $Li_6PS_5Cl$ (LPSC) was synthesized as described in detail elsewhere[41]. Appropriate amounts of $Li_2S$ (99.9%, Alfa Aesar), $P_2S_5$ (99%, Sigma-Aldrich) and LiCl (99.0%, Sigma-Aldrich) were ball-milled at 110 rpm for 1 h under argon atmosphere. The mixture was then transferred to quartz tubes and annealed at 550 °C for 10 h in order to get the pure phase of the argyrodite $Li_6PS_5Cl$.

**Solid-state battery assembly and electrochemical cycling:**

The electrode mixture was prepared by ball milling argyrodite with carbon (Super P, TIMCAL) and carbon nanofibres (Sigma Aldrich) in a weight ratio of 0.70 : 0.15 : 0.15 for 6 h at 450 rpm in a $ZrO_2$ coated stainless steel jar with 8 $ZrO_2$ balls. The solid electrolyte and electrodes were then cold pressed under 4 tons/cm$^2$ in a solid-state cell. In a cell, 10 mg of LPSC-C electrode was used and pressed against 140 mg of electrolyte[41,42]. Cycling was performed in an argon filled glove box, in order to avoid reactions with oxygen and moisture. The ASSBs were cycled galvanostatically with a current density of 5.5 mA/cm$^2$ within a voltage window of 0 – 3 V vs. Li/In for In|LPSC|LPSC-C, 0 – 1 V vs. Li/In for Li-In|LPSC|LPSC-C and 0 – 2.5 V for LPSC-C|LPSC|LPSC-C respectively. To evaluate the practical electrochemical stability window more accurately, the differential capacity is determined from the 1$^{st}$ charge of the In|LPSC|LPSC-C battery and from the 1$^{st}$ discharge of the Li-In|LPSC|LPSC-C battery. Often cyclic voltammetry (CV) is used to determine the experimental stability window. However, the relatively short exposure time to the decomposition potentials in combination with the typically sluggish decomposition reactions make it challenging to evaluate the electrochemical stability window with CV cycling. In contrast, the differential capacity, determined from the slow galvanostatic charge and discharge profiles of individual oxidation and reduction processes is effective in determining the practical electrochemical stability window, in particular when the solid electrolyte is used as an active electrode material. To measure the oxidative and reductive stability of LLZO and LAGP an NMP (Sigma Aldrich) based electrode slurry was prepared by ball-milling active material (LLZO Ta-doped, D50 = 400 – 600 nm, Ampcera™, LAGP, Ampcera™), with carbon black (Super P, TIMCAL), PVDF binder (polyvinylidene fluoride, Solef® PVDF, Solvay ) in weight ratio 0.4 : 0.5 : 0.1 for 90 min at 250 rpm in $ZrO_2$ coated stainless steel jar with 8 $ZrO_2$ balls. A blank test was prepared using carbon black (Super P, TIMCAL) as active material and PVDF as a binder in the weight ratio 0.9 : 0.1 to result in the same carbon black loading as the LLZO and LAGP electrodes. The slurry was casted on Al foil with a thickness of 100 µm and dried at 60°C in vacuum oven for 12h. The loading of the LLZO, LAGP and carbon electrodes was 1.6 mg/cm$^2$, 1.0 mg/cm$^2$ and 0.6 mg/cm$^2$ respectively. The coin cells were assembled in an argon filled glove box, in order to avoid reactions with oxygen and moisture (< 0.1 ppm $O_2$ and < 2 ppm $H_2O$) using both a polymer (Celgard 2250) and glass fiber (Whatman) separator and lithium metal as a counter electrode (Sigma Aldrich), which is washed with dimethyl carbonate (DMC) to remove the oxide layer and 400 µl of 1.0 M $LiPF_6$ in 1:1 v/v ethylene carbonate (EC) and diethyl carbonate (DEC) (<15 ppm $H_2O$, Sigma Aldrich) was added as an electrolyte for wetting both working and counter electrode surfaces. Galvanostatic oxidation was performed with cut-off voltage of 4.5 V (vs. Li/Li$^+$) for first oxidation (LLZO-C) and 1.5 V (vs. Li/Li$^+$) for first reduction (LAGP-C) with 12 hours of rest and charge/discharge current of 7.0 µA. Comparison of the galvanostatic oxidation and reduction of the LLZO-C and LAGP-C electrodes and blank electrode are provided in Supplementary Fig. 8a,b. With the solid electrolyte – carbon mixtures, very large interface areas are achieved (for the current particle sizes >>1 m$^2$) making the effective current densities at least 4 orders of magnitude lower than the current densities based on the electrode diameter.

**X-Ray Diffraction:**

In order to identify the crystalline phases of the prepared materials, powder XRD patterns were collected in the 2θ range of 10–120° using Cu Kα X-rays (1.5406 Å at 45 kV and 40 mA) on an X'Pert Pro X-ray diffractometer (PANalytical). The samples were tested in an airtight sample holder, filled with argon, to prevent exposure to oxygen and moisture.

**Solid-state NMR:**

Solid-state (NMR) measurements were performed using a Bruker Ascend 500 MHz spectrometer equipped with two and three channel 4.0 mm and 3.2 mm Magnetic Angle Spinning (MAS) probes respectively. The operating frequencies for $^{31}$P and $^{6}$Li were 202.47 and 73.60 MHz respectively, and all measurements were performed within a spinning speed range of 8 to 23 kHz and π/2 pulse lengths of 4 – 5 μs were determined for $^{6}$Li and $^{31}$P. The chemical shifts of $^{6}$Li spectra were referenced with respect to a 0.1M LiCl solution, and $^{31}$P spectra with respect to an 85% $H_3PO_4$ solution. Based on the $T_1$ relaxation time, recycle delays of 5-10000 s were utilized collecting between 128 and 11264 scans for each sample.

**Computational details :**

To determine the energy properties of crystalline phases, Density Functional Theory (DFT) relaxations were performed with the Vienna Ab-Initio Simulation Package (VASP)[43]. The PBE exchange correlation function of Perdew et al. was implemented[44], and core electrons were probed with the projected-augmented wave approach (PAW)[45]. A cut-off value of 280 eV and a 4 x 4 x 4 k-point mesh were used. For the argyrodite, the following reaction is considered:

$$Li_6PS_5Cl \rightarrow Li_xPS_5Cl + (6-x)Li \quad (1)$$

If x < 6 Li Li$_6$PS$_5$Cl is oxidized, if x > 6 Li$_6$PS$_5$Cl is reduced. Then, by calculating the energies on both sides of the reaction and taking the electrochemical potential of Li into account,

$$\bar{\mu}_{Li} = \mu_{Li} - \phi \quad (2)$$

with $\bar{\mu}_{Li}$ the electrochemical potential of Li, $\mu_{Li}$ the chemical potential of Li, and $\phi$ the electrical potential. Therefore, the average electrical potential at which oxidation/reduction takes place can be determined by:

$$\bar{\phi} = -\frac{E(Li_6PS_5Cl) - E(Li_xPS_5Cl) - (6-x)E(Li)}{6-x} \quad (3)$$

Where $E(Li_xPS_5Cl)$ represents the composition of the most stable configurations on the convex hull. DFT based Molecular Dynamics (MD) simulations were performed using the same cutoff value as in DFT simulations. The ab initio MD simulations were executed in the NVT ensemble, where the temperature scales every 1000 time steps. The simulations use periodic boundary conditions with time steps of 2 fs, the total time of the MD simulations being 100 ps. The number of k-points was reduced from 4 x 4 x 4 used in the DFT simulations to 1 x 1 x 1 for the MD simulations. The lattice parameters and positions of all atoms were allowed to relax during relaxation.

The argyrodite structure was obtained from previous work[27]. There the Cl-S disorder over the 4a and 4c sites was investigated, and the thermodynamically most favourable configuration was obtained. Note that the Cl-S disorder is kept constant in the presented convex hull and thus the oxidation and reduction voltages are not affected. For determination of configurations as a function of the Li-concentration, 10000 structures are created by placing the appropriate amount of Li-ions randomly at the 48h-positions. To quickly scan these for possible low-energy structures only the electrostatic energies in these structures were calculated, using the Undamped Shifted Force method with a cut-off radius of 15 Angstrom[46]. For the 20 lowest energy configurations of the electrostatic calculations the structure was optimised and the energy was calculated using VASP. For $Li_xPS_5Cl$ 12 ≤ x ≤ 15 extra Li atoms are inserted on the 16e position.

All DFT calculations are performed on charge-neutral cells, thus taking into account the true oxidation and reduction of solid electrolytes, and thus behaving similar to an electrode material. The formation energies of the thermodynamic decomposition products are taken from the Materials Project database[47]. The structure of LLZO is obtained from the Materials Project database[47]. For LLZO a 1 x 1 x 1 k-point mesh was used with a cut-off value of 500 eV. The structure of LAGP is taken form literature[48] and was relaxed using a 3 x 3 x 1 k-point mesh with a cut-off value of 500 eV.

**Acknowledgements**

The authors thank Kees Goubitz, Michel Steenvoorden, and Frans Ooms for their assistance with experiments and Carla Robledo for her assistance with the schematic graphic. Financial support is greatly acknowledged from the Netherlands Organization for Scientific Research (NWO) under the VICI grant nr. 16122, from the eScience Centre and NWO under the joint CSER and eScience programme for Energy Research grant nr. 680.91.087 and from the Advanced Dutch Energy Materials (ADEM) program of the Dutch Ministry of Economic Affairs, Agriculture and Innovation.

## Author Contributions

TS, AV and NdK did the DFT simulations and TS and AV analysed data. CY and CW synthesized the solid electrolyte. TS, CW, VA did the electrochemical measurements, and TS and VA analysed the data. TS, VA and CW did the XRD measurements and VA, TS and SG analysed the data. VA measured and analysed the NMR data. CY, EvdM, YX and JH did preliminary measurements. TH, IK and EMK contributed to the discussion of results. MW supervised the project, MW, SG and CY designed the research and TS, VA, SG and MW wrote the manuscript.

## Competing Interests

The authors declare no competing financial interests.